\documentstyle[prl,aps,epsf,multicol]{revtex}
\begin{document}
\draft

\def\i{\imath\,}
\def\ih{\frac{\imath}{2}\,}
\def\undertext#1{\vtop{\hbox{#1}\kern 1pt \hrule}}
\def\ra{\rightarrow}
\def\lfa{\leftarrow}
\def\Ra{\Rightarrow}
\def\lra{\longrightarrow}
\def\ler{\leftrightarrow}
\def\lrb#1{\left(#1\right)}
\def\O#1{O\left(#1\right)}
\def\EV#1{\left\langle#1\right\rangle}
\def\tr{\hbox{tr}\,}
\def\trb#1{\tr\lrb{#1}}
\def\dd#1{\frac{d}{d#1}}
\def\dbyd#1#2{\frac{d#1}{d#2}}
\def\pp#1{\frac{\partial}{\partial#1}}
\def\pbyp#1#2{\frac{\partial#1}{\partial#2}} 
\def\pd#1{\partial_{#1}}
\def\br{\\ \nonumber & &}
\def\brr{\right. \\ \nonumber & &\left.}
\def\inv#1{\frac{1}{#1}}
\def\be{\begin{equation}}
\def\ee{\end{equation}}
\def\bea{\begin{eqnarray}}
\def\eea{\end{eqnarray}}
\def\ct#1{\cite{#1}}
\def\rf#1{(\ref{#1})}
\def\EXP#1{\exp\left(#1\right)} 
\def\INT#1#2{\int_{#1}^{#2}} 
\def\LHS{left-hand side }
\def\RHS{right-hand side }
\def\COM#1#2{\left\lbrack #1\,,\,#2\right\rbrack}
\def\AC#1#2{\left\lbrace #1\,,\,#2\right\rbrace}

\title{Quasiparticle localization in superconductors with spin-orbit scattering}
\author{T. Senthil and Matthew P.A. Fisher}
\address{
Institute for Theoretical Physics, University of California,
Santa Barbara, CA 93106--4030
}

\date{\today}
\maketitle

\begin{abstract}
We develop a theory of quasiparticle localization in superconductors in situations 
without spin rotation invariance. We discuss the existence, and properties
of superconducting phases with localized/delocalized quasiparticle excitations in 
such systems in various dimensionalities. Implications for a variety of experimental systems,
and to the properties of random Ising models in two dimensions, are briefly 
discussed. 
 
\end{abstract}
\vspace{0.15cm}


\begin{multicols}{2}
\narrowtext 
\section{Introduction}

A powerful probe of the properties of a superconductor is 
obtained by studying the low temperature dynamics of the quasiparticles.
In this context, we proposed\cite{smitha} that all 
ground state phases of disordered superconductors
can be characterized,
at zero temperature, by their quasiparticle transport properties. 
The two general possibilities are that the quasiparticle excitations may be delocalized,
analogous to a metal, or be localized analogous to an insulator.
Previous 
papers\cite{short,dos,carlos,sqh} have developed a theory
of localization of quasiparticles in superconductors
in situations with spin rotation invariance. In this paper, 
we consider the case where the spin is not conserved. 
This may happen, for instance, in a singlet superconductor
in the presence of spin-orbit scattering. 
Another example is 
provided by a triplet superconductor where the 
quasiparticles can exchange spin with the condensate, and 
hence, do not have conserved spin.
Indeed, a number of  
superconducting systems, such as for instance the heavy 
fermion superconductors, 
are both strongly disordered and have strong spin-orbit scattering (and perhaps
even triplet pairing). Thus, in order to
understand the possibility of quasiparticle localization in such systems, it is necessary 
to develop a theory that includes spin-orbit scattering in an essential way.   
Besides, by analogy with
what happens in normal metals,  spin-orbit scattering is 
expected to have profound effects on localization phenomena.  

Because the quasiparticle
charge density is also not conserved in a
superconductor, the only conserved quantity 
carried by the quasiparticles (at low
energies) is the energy density itself. 
In the presence of impurity scattering, the quasiparticle
charge and spin densities in such a superconductor
do not diffuse, as they are not conserved. Energy diffusion
is possible, though. The corresponding transport quantity is the quasiparticle thermal
conductivity. 

From a theoretical point of view,  
quasiparticles in a superconductor with non-conserved spin are 
more appropriately thought of as real (Majorana) fermions. Thus the problem we consider here
is one of localization of Majorana fermions. While localization issues of 
complex (conventional) fermions have been explored in considerable detail, 
surprisingly, there has been very little theoretical work on 
corresponding issues for Majorana fermions. As we argue, the superconductor
with non-conserved spin provides a natural experimental realization of 
such a system. We examine the possible
phases (as characterized by quasiparticle transport) and 
the associated phase transitions. It is of interest to distinguish between
situations with and without time reversal symmetry ${\cal T}$, 
and we will consider each separately. (In the notation of Ref. \cite{AZ}, these 
correspond to Class DIII and Class D respectively).

While the superconductor is our primary motivation, we note also that 
Majorana fermions arise in other contexts as well. A well-known example is the 
two dimensional Ising model. There have been several studies\cite{RBIM} of the properties of the
two dimensional Ising model in the presence of randomness in the bond strengths, though
there are still several poorly understood issues. The implications of this work for that 
problem will be considered briefly towards the end of the paper.

\begin{figure}
\epsfxsize=3.5in
\centerline{\epsffile{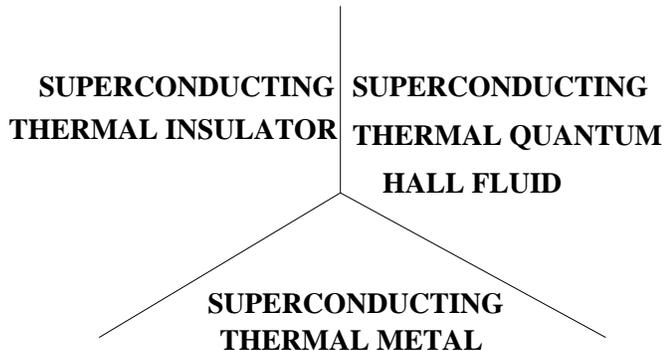}}
\vspace{0.15in}
\caption{Schematic zero temperature phase diagram for the two dimensional
superconductor in the absence of both spin rotation and time reversal
invariances.}
\vspace{0.15in}
\label{major2d}
\end{figure}

We first show the existence, in two dimensions, of stable ``metallic''
and ``insulating'' phases inside the superconductor with delocalized
and localized quasiparticle excitations respectively. The stability of the
``metallic" phase in two dimensions has already been alluded to in Ref. \cite{carlos}.
We emphasize that both
phases are superconducting - they are distinguished by the nature of thermal transport
due to the quasiparticles. These two phases are separated by a phase transition
which is a novel ``metal-insulator" transition inside a superconductor. In Ref. \cite{smitha},
we discussed possible experimental realizations of such phase transitions. The 
universal critical properties of this transition depend, of course, 
on whether time reversal is a good symmetry 
or not. In ${\cal T}$-non-invariant systems, the insulating phases may be further 
characterized in terms of their values of the Hall thermal conductance. The dimensionless
ratio $\frac{3h\kappa_{xy}}{ \pi^2 k_B^2 T}$ is quantized in units of $1/2$ . Phases with different values of this 
quantized Hall thermal conductance are topologically distinct, and are separated by phase
transitions. In Figure \ref{major2d}, we show a schematic phase diagram. Note that,
in the case with no ${\cal T}$, apart from the ``metal-insulator" transition, there are 
also transitions between the ``metal" and the ``quantum Hall" phase, and from 
the insulator with $\kappa_{xy}/T =0$ to the ``quantum Hall" phase. Further,
there is a multicritical point, where all three phases come together. 
In three dimensional systems, ``metallic" and ``insulating" phases are again possible
with the transition between the two being in a new universality class for 
localization. In the rest of the paper, we will 
substantiate these claims, and analyse the properties 
of each phase in further detail.

\section{Models and general formalism}
Consider a general lattice Hamiltonian for the quasiparticles
in a superconductor with strong spin-orbit scattering:
\begin{equation}
\label{hso}
{\cal H} = \sum_{ij} t^{\alpha \beta}_{ij}c^{\dagger}_{i\alpha}c_{j\beta}
+ \Delta^{\alpha \beta}_{ij} c^{\dagger}_{i\alpha}c^{\dagger}_{j\beta} + h.c  .
\end{equation}
Here $i,j$ refer to the sites of a lattice, 
and $\alpha, \beta$ are spin indices. We assume that $t$ and $\Delta$
are both short-ranged in space. 

We will focus on two cases - with and without time reversal
symmetry (${\cal T}$).
If present, time reversal is imposed  
through an  antiunitary ``time-reversal" operator ${\cal T}$ 
which transforms the $c$ operators as:
\begin{equation}
{\cal T}\left[\begin{array}{c}c_{i\uparrow} \\
c_{i\downarrow} \end{array}\right] = \left[\begin{array}{c}c_{i\downarrow} \\
                                                          -c_{i\uparrow} \end{array}\right]
=i\sigma_y c   . 
\end{equation}
Note that ${\cal T} \bbox{\sigma}^* {\cal T}^{-1} = - \bbox{\sigma}$
(with $\bbox{\sigma}$ a vector of Pauli matrices) so that
the electron spin is odd under time reversal.  With time reversal invariance
present we require that
\begin{equation}
\label{Tinv1}
{\cal T}{\cal H}{\cal T}^{-1} = {\cal H}   .
\end{equation}

Note that ${\cal H}$ has no special symmetries 
(other than possibly ${\cal T}$). 
In particular, neither the charge nor spin is conserved. It is
convenient then to work with Majorana fermions $\eta_{1i\alpha}, \eta_{2i\alpha}$
defined through
\begin{equation}
c_{i\alpha} = \frac{1}{\sqrt{2}}\left(\eta_{i1\alpha} + 
i \eta_{i2\alpha}\right)  .
\end{equation}
The Hamiltonian
when expressed in terms of the $\eta$ Fermions takes the form
\begin{equation}
\label{hsom}
{\cal H} = \eta^T H \eta  ,
\end{equation}
with $\eta = \eta_{ia\alpha}$, ($a, b = 1,2$) and $H= H_{ij}^{ab, \alpha,\beta}$ is a matrix 
in $(ij), (\alpha \beta), ab$ space. By definition, $H$ is
hermitian and moreover $H^T = -H$, so that $H$ is pure imaginary. 
Thus the problem of quasiparticle localization in a superconductor 
with spin-orbit scattering is, in essence, one of localization of 
Majorana fermions. In particular, in the case where even ${\cal T}$ is not a good symmetry,
the Hamiltonian Eqn. \ref{hsom} is the most general one describing
non-interacting Majorana fermions in a disordered system.  

Time reversal symmetry is easily imposed on the Majorana Hamiltonian Eqn. \ref{hsom}.
Under the action of the antiunitary operator ${\cal T}$, it is readily seen that
\begin{equation}
\label{Teta}
{\cal T}\eta_{ia\alpha} = i(\sigma_y)_{\alpha \beta} (\tau_z)_{ab} \eta_{ib\beta}  .
\end{equation}
Here $a,b = 1,2$ and $\vec \tau$ is a Pauli matrix in $ab$ space. 
The time reversal
invariance condition Eqn. \ref{Tinv1} combined with Eqn. \ref{Teta} implies
\begin{equation}\label{TinvH}
\sigma_y\tau_z H^* \sigma_y \tau_z = H  .
\end{equation}
 
We note in passing that for a superconductor of spin {\it polarized}
Fermions (appropriate, say, in the $A-1$ phase of superfluid
$3-He$) in the absence of time reveral invariance,
the general Hamiltonian can still be written in the form
${\cal H} = \eta^T H \eta$, with $H$ an antisymmetric and pure imaginary
matrix in position ($i,j$) and ``particle-hole" ($ab$) space.
But for spinless quasiparticles
the condition for time reversal invariance
is different, leading to different symmetries.
Specifically, with time reversal invariance the lattice Hamiltonian
depends on a {\it real} symmetric hopping matrix, $t_{ij}$,
and a {\it real} antisymmetric gap matrix, $\Delta_{ij}$.
In this case, when re-expressed in terms of Majorana fermions,
the Hamiltonian becomes ${\cal H} = \eta^T H \eta$
with $H = t \tau_y + i \Delta \tau_x$.
This Hamiltonian matrix 
can equivalently be expressed as $H= i A \tau^+ + c.c.$,
with $\tau^+ = \tau_x + i \tau_y$ and $A= t+\Delta$
an {\it arbitrary} real matrix.  
This 
``off-diagonal" form is very different than the
two cases with spinful electrons, and in fact
belongs to a different symmetry class - 
a class studied by Gade and Wegner\cite{Gade}.
Henceforth, we focus exclusively on the spinful case.

We are interested in understanding the nature of energy transport (in both spinful cases, with and 
without ${\cal T}$)
by the excitations described by the Hamiltonian. For this
purpose,
it is actually convenient to adopt the following trick. We 
consider two
identical copies of the system. To describe the two copies, we introduce
two Majorana fields $\eta$ and $\zeta$ and consider
\begin{equation}
\tilde {\cal H} = \eta^T H \eta + \zeta^T H \zeta  .
\end{equation}
It is now possible to combine the $\eta$ and $\zeta$ fields
into a single complex fermion $f$:
\begin{eqnarray}
f & = & \frac{1}{\sqrt{2}}(\eta + i \zeta)  ,\\
f^{\dagger} & = & \frac{1}{\sqrt{2}}(\eta - i\zeta) .
\end{eqnarray}  
Then, we have
\begin{equation}
\label{hsotf}
\tilde{\cal H} = f^{\dagger}Hf  .
\end{equation}
Note that the number of $f$ particles is
conserved. It is then possible to consider transport of this 
conserved $f$-number density. This may be quantified by a
conductivity $\sigma_f$. As the Hamiltonian $\tilde{\cal H}$
describes non-interacting $f$-particles, the thermal conductivity $\kappa_f$
of the $f$ particles is related to $\sigma_f$ by a Weiedemann-Franz
law (as $T \rightarrow 0$):
\begin{equation}
\label{WF}
\frac{\kappa_f}{T\sigma_f} = \frac{\pi^2 k_B}{3}  .
\end{equation}
Since the Hamiltonian $\tilde{\cal H}$ represents just two identical copies
of the original system described by ${\cal H}$, it is clear that the thermal
conductivity ($\kappa$) of the $\eta$ particles is exactly half that of the 
$f$-particles:
\begin{equation}
\kappa = \frac{\kappa_f}{2}  .
\end{equation}
Combining this with Eqn. \ref{WF}, we see that
calculation of $\kappa$ is reduced to calculation of 
$\sigma_f$. Similarly, all the thermodynamic properties of ${\cal H}$
may be obtained by halving the corresponding property calculated 
with $\tilde{\cal H}$. 

It is convenient to define the $f$-Green's function
\begin{equation}
G_{ij}^{ab, \alpha\beta}(E) = \langle ia\alpha | \frac{1}{E - H + i\epsilon}|jb\beta \rangle  ,
\end{equation}
where $\epsilon$ is a positive infinitesimal and $E$ is the energy (measured
from the Fermi energy). The condition $H^* = -H$ immediately implies that
\begin{equation}
\label{Gsymm}
G^*(E) = -G(-E),
\end{equation}
where we have suprressed all the indices on $G$. In particular, for states at the 
Fermi energy, we have
\begin{equation}
\label{Gsymmo}
G^*(E = 0) = -G(E = 0)  ,
\end{equation}
so that $G(E=0)$ is pure imaginary.
This 
Green's function can be convenintly expressed, as usual, as a functional integral over 
Grasmann variables $f_{ia\alpha}, \bar{f}_{ia\alpha}$ with the action
\begin{equation}
S = i\bar{f}(H - i\epsilon)f .
\end{equation}
The symmetry Eqn. \ref{Gsymmo} implies that this same generating functional
can be used to calculate transport quantities\cite{dos}.

\section{Systems with time reversal symmetry}
\label{Tso}
The discussion has so far been completely general. We now specialize 
to the case with ${\cal T}$ symmetry.
\subsection{``Metallic" Phase}
\label{Tmet}
 We first consider 
the situation in which the disorder is weak.
We assume that there is a finite,
non-zero mean free path $l_e$ set by the impurity strength in such a manner that
the $f$-particle motion is diffusive on larger scales. 
(In terms of the original physical system, this corresponds to diffusion
of energy). 
We are interested
in describing the effects of quantum interference on this diffusive motion. 
We assume also that, in the absence of quantum interference effects, the density
of states at the Fermi energy of this diffusive system is finite and non-zero.
In that case, it is possible to follow standard techniques to derive a 
replica non-linear sigma model field theory to describe the physics at 
length scales large compared to the mean free path $l_e$. As the procedure is
sufficiently well-known, we merely state the results. The field theory is 
described by the action
\begin{equation}
\label{sotft}
{\cal S} = \int d^d x \frac{1}{2g} Tr\left (\nabla O)^T (\nabla O) \right)
-\epsilon Tr (O + O^T)  ,
\end{equation}
where $O(x) \in O(2n)$ is a $2n \times 2n$ orthogonal matrix-valued field with 
$n$ the number of replicas.
When $\epsilon = 0$ the action has a global 
$O(2n) \times O(2n)$ symmetry,
$O \rightarrow AOB$ with $A$ and $B$ orthogonal matrices,
which is broken down to the diagonal $O(2n)$ ($A^T = B$)
by the $\epsilon$ term.
The coupling constant $g$ is inversely proportional to the 
bare $f$-conductivity 
$\sigma_f^0$ (i.e the conductivity on the scale of the mean free path). 
The density of quasiparticle states at the Fermi energy is given by
\begin{equation}
\rho = \frac{\rho_0}{4n}\langle Tr(O + O^T) \rangle.
\end{equation}

Consider a renormalization group transformation
where short distance fluctuations are integrated out, and the coordinate $x$ is
rescaled as $x \rightarrow x' = x e^{-l}$.
The leading quantum interference corrections to diffusion can now
be obtained from the known\cite{Zinn-Justin} 
perturbative $\beta$ function of this field theory
in the replica limit. The result, in two dimensions, is
\begin{equation}
\frac{dg}{dl} = -\frac{g^2}{4\pi}  .
\end{equation}

Note that $g$ decreases as  $l$ is increased. Thus the 
perturbation theory (in powers of $g$) gets better at large length scales.
For a system of size $L$, at zero temperature, we may integrate the flow
equation upto a scale $l^*$ given by $\l_e e^{l^*} \sim L$ to get
\begin{equation}
g(L) = \frac{g_0}{1 + \frac{g_0}{4\pi}\ln\left(\frac{L}{l_e}\right)}  ,
\end{equation}
where $g_0$ is the bare value of $g$. For large $L$, this therefore
gives
\begin{equation}
g(L) \approx \frac{4\pi}{\ln\left(\frac{L}{l_e}\right)}  .
\end{equation}
Thus $g(L)$ goes to zero logarithmically with the system size. As $\sigma_f$
is inversely proportional to $g$, it follows that $\sigma_f$ {\em diverges}
logarithmically with the system size. At finite temperature, in an infinite system,
it is natural to expect that the quantum interference effects will be cut-off
at a finite dephasing length scale $L_{\phi} \sim T^{-p}$ due to interaction
effects not included in the model. We therefore have
\begin{equation}
\sigma_f \sim \ln\left(\frac{1}{T}\right)  ,
\end{equation}
at the lowest temperatures. As this also determines the 
thermal conductivity $\kappa(T)$ of the original system, we have
\begin{equation}
\frac{\kappa}{T} \sim \ln\left(\frac{1}{T}\right)  .
\end{equation}

The considerations above establish the existence of a ``metallic" phase with
delocalized quasiparticle excitations in two 
dimensions in the model Hamiltonian 
Eqn. \ref{hso} describing the superconductor in the presence of 
time reversal invariance, but no spin rotation invariance. 
In striking contrast to normal metals, quantum interference effects
also lead to singular corrections to the density of states in a 
superconductor\cite{dos}. In our previous work\cite{dos}, we 
demonstrated this in the spin-rotation invariant cases. 
We now show that the density of states is {\em enhanced} 
in the situation considered
in this paper. In particular, we show that in
two-dimensions it actually {\em diverges}
in the thermodynamic limit. 

To see this, consider the action Eqn. \ref{sotft} at finite $\epsilon$.
The density of states is obtained from
\be
{ {\rho }   \over  {\rho_0} } = -\frac{1}{4n}\left[\frac{\partial {\cal F}}{\partial \epsilon}\right]_{\epsilon = 0}  ,
\ee
where ${\cal F}$ is the free energy density defined through
\be
\EXP{-L^d{\cal F}} = \int{\cal D}O ~ \EXP{-{\cal S}}.
\ee
The flow of $\epsilon$ under the renormalization group may be obtained straightforwardly. To leading order in $g$ the result
is
\be
\dbyd{\epsilon}{l} = \lrb{2 + \frac{g}{8\pi}}\epsilon.
\ee
This is readily integrated to give
\be
\epsilon(l) = \epsilon(0)\EXP{2l + {1 \over 8  \pi} \INT{0}{l}g(l') dl'}  .
\ee
The free energy density scales according to
\be
{\cal F}(g(0), \epsilon(0)) = e^{-2l}{\cal F} \lrb{g(l), \epsilon(l)}  ,
\ee
where we have specialized to two-dimensions.
For a system of size $L$, run the RG till a scale $l^*$ such that $l_e e^{l^*} \approx L$.
The density of states is then
\bea
\rho(L) & = & -\frac{e^{-2l^*}}{4n}\pbyp{{\cal F}\lrb{g(l^*), \epsilon(l^*)}}{\epsilon(0)} \nonumber \\
& = & -\EXP{\frac{1}{8\pi}\INT{0}{l^*} dl' g(l')}  \frac{1}{4n}
\pbyp{{\cal F}\lrb{g(l^*), \epsilon(l^*)}}{\epsilon(l^*)}
\eea
After scaling out to $l^*$ the mean free path
is comparable to the system size, so that
$- \frac{1}{4n}
\pbyp{{\cal F}\lrb{g(l^*), \epsilon(l^*)}}{\epsilon(l^*)}  \ra const.$. Therefore,
\be
\rho(L) \sim \EXP{\frac{1}{8\pi}\INT{0}{l^*} dl' g(l')}  .
\ee
For large $l$, $g(l) \approx \frac{4\pi}{l}$. Thus, we have 
\be
\rho(L) \sim \sqrt{\ln \frac{L}{l_e}}  .  
\ee
Thus the density of states at the Fermi energy diverges. The behaviour of the density of states as 
a function of energy in an infinite system may also be found similarly (or simply guessed 
from the equation above) to be
\be
\rho(E) \sim \sqrt{\ln \frac{1}{E}}   .
\ee 

The Einstein relation can now be used to infer that the heat diffusion constant $D$ also diverges
on approaching the Fermi energy as
\be
D(E) \sim \sqrt{\ln \inv{E}}  .
\ee
Thus the ``metallic" phase has an infinite heat diffusion constant at zero temperature in two dimensions.

The divergence of the density of states has obvious consequences for the low temperature thermodynamic
properties of this phase. For instance, the specific heat behaves as
\be
C(T) \sim T\sqrt{ \ln \inv{T}}   .
\ee

In three dimensional systems, the stability of the ``metallic'' phase can be established
by simple power-counting arguments. Quantum interference effects are then irrelevant at long length scales.
The thermal conductivity then goes to zero linearly with the temperature at low $T$:
\be
\frac{\kappa_{3D}}{T} \ra const  .
\ee
The density of states
at the Fermi energy is finite, and non-zero. However, quantum interference effects do lead to a singular
$\sqrt{|E|}$ cusp in the density of states as a function of the energy (See Ref. \ct{smitha} for an
analogous discussion in the superconductor with conserved spin) so that
\be
\rho(E) - \rho(E = 0) \sim - \sqrt{|E|}  .
\ee
Note that the density of states increases with decreasing energy as also happens in $d = 2$. 
  
\subsection{``Insulating" phase}
\label{Tins}
For strong disorder, it is possible to have a phase with localized quasiparticle excitations
in any finite dimension. 
In the terminology of Ref. \ct{smitha}, this is a superconducting `` thermal insulator''. 
In this phase, the ratio $\frac{\kappa}{T}$ goes to zero with the temperature. 
The density of quasiparticle states also goes to zero at the Fermi energy. To see 
this, consider the lattice Hamiltonian Eqn. \ref{hsotf} in the extreme limit where
the $f$ particles are localized to a single site. The Hamiltonian for a single site is constrained
to be of the form $H = a \sigma_y \tau_x + b \tau_y$ with $a,b$ real. 
This has two eigenvalues given by $ \pm \sqrt{a^2 +b^2}$. Consider now the case where the 
distribution of $a,b$ has finite, non-zero weight at $a=b=0$. Then, the density of states $\rho(E)$
averaged over the disorder vanishes as $|E|$. If the distribution has vanishing weight at $a = b = 0$, 
then $\rho(E)$ vanishes faster than linearly. Including hopping between the sites should not 
change this result so long as we are in the localized phase. 
(As with the superconductors with conserved spin, 
having a finite density of states requires a diverging weight at $a=b=0$ which is presumably unphysical,
and non-generic). We thus conclude that the density of quasiparticle states vanishes, at least as fast as 
$|E|$ in the localized phase.

\section{Time reversal broken systems}
We now move on to systems without time reversal symmetry. As mentioned earlier,
this corresponds to studying the general localization properties of a 
non-interacting disordered system of Majorana fermions. It is therefore
appropriate to call the phase with extended states a ``Majorana metal'', and the phase with
localized states a ``Majorana insulator''. 
\subsection{Majorana metal}
To address issues such as the stability and properties of the Majorana metal, it is
useful to think in terms of a replica sigma model field theory which permits a systematic
study of quantum interference corrections to diffusive energy transport. This is 
readily done using standard techniques. The result is
\be
\label{sd}
{\cal S} = \int d^2x \frac{1}{2g} Tr \lrb{\partial Q}^2 - \epsilon Tr \lrb{\sigma_yQ},
\ee
where $Q = V^T \sigma_y V$ is a $2n \times 2n$ matrix, $V$ is an $O(2n)$ matrix and $n$ is again
the replica index. 
When $\epsilon =0$, the action is invariant under the group $O(2n)$,
$Q \rightarrow W^T Q W$, with $W$ an orthogonal matrix.
With non-zero $\epsilon$, invariance of the action
requires $W = exp(iG)$ with
$G$ pure imaginary and antisymmetric of the form,
$W = S\sigma_y + iA$.  Here $S$ is an $n$ by $n$ real symmetric matrix and
$A$ is $n$ by $n$ real antisymmetric.
Since $S+iA$ is $n$ by $n$ Hermitian,
the group has evidently been broken down to $U(n)$
by the $\epsilon$ term. Thus ${\cal S}$ describes a non-linear sigma model on the manifold
$O(2n)/U(n)$. The coupling constant $g$ is again inversely proportional to the 
(longitudinal) $f$-particle conductance. The density of states of the quasiparticles
may be obtained from
\be
\rho = \inv{2n}\EV{Tr\lrb{\sigma_yQ}}  .
\ee

Actually, on symmetry grounds, there is a topological term allowed in the action. This 
follows from the observation that $\Pi_2\lrb{O(2n)/U(n)} = Z$ is non-trivial. This will be 
important to understand the behaviour of the thermal Hall conductivity later.

The perturbative RG flows of the couplings $g$ and $\epsilon$ in two dimensions in the replica limit are
\bea
\dbyd{g}{l} & = &  -\frac{g^2}{8\pi}   , \\
\dbyd{\epsilon}{l} & = & \lrb{2+ \frac{g}{8\pi}}\epsilon  .
\eea
We may use these to extract the physical properties of the Majorana metal phase in 
exactly the same manner as in the previous section. We therefore simply state the results. 
First note that the (marginal) irrelevance 
of the coupling $g$ implies the stability of the Majorana metal in $d = 2$. 
As the temperature goes to zero, 
the (longitudinal) thermal conductivity $\kappa_{xx}$  in a finite system
of size $L$ behaves as
\be
\frac{\kappa_{xx}}{T} \sim \ln \lrb{\frac{L}{l_e}}  .
\ee
In an infinite system at finite $T$, where the quantum interference is cut-off by dephasing due to
interaction effects, we have
\be
\frac{\kappa_{xx}}{T} \sim \ln\left(\frac{1}{T}\right)  .
\ee
Similarly, the density of states at the Fermi energy now diverges logarithmically with the system size:
\be
\rho(L) \sim \ln \frac{L}{l_e}  .
\ee
As a function of energy in an infinite system, we have
\be
\rho(E) \sim \ln \inv{E}  .
\ee
This leads, for instance, to a specific heat which depends on temperature as
\be
\frac{C(T)}{T} \sim \ln \inv{T}   .
\ee
Since $\kappa_{xx}/T$ and the density of states diverge
in the same manner with energy,  determinining the behavior of
the thermal diffusion coefficient would require going
to second order in the perturbative RG.  The diffusion coefficient
could then also diverge,
but not more rapidly than
double logarithmically with energy.

In three dimensions, the behaviour of the superconducting ``metal" phase is qualitatively the same
irresepective of whether time reversal symmetry is present or not. Therefore, the discussion in the 
previous section of the three dimensional case applies here as well.
 
\subsection{Majorana insulator}
At strong disorder, in any dimension, it is possible to find phases where the quasiparticle excitations are
localized. The (longitudinal) quasiparticle heat conductivity $\kappa_{xx}$ goes to zero rapidly with the temperature,
in such a phase. We will call this the Majorana insulator, as this corresponds to a localized phase of 
Majorana fermions. The density of states of the Majorana insulator may be found by an argument analogous 
to the one in Section \ref{Tins}. In brief, we consider a single site Hamiltonian for two Majorana species 
(corresponding
to a single spinless complex Fermion)
which is constrained to be of the form $H = a\tau_y$ with $a$ real
and random. If the distribution of $a$ has a finite,
non-zero weight at $a = 0$, then there is a
finite density of states at zero energy. We again expect this to hold 
through out the localized phase, so that the density of states 
at the Fermi energy can be non-zero. Note that, like in a conventional insulator in non-superconducting systems, a 
density of states that vanishes at the Fermi energy is also possible.

\subsection{Majorana quantum Hall phase}
In the absence of time reversal symmetry, there is potentially another transport property which can be used to 
distinguish zero temperature phases: the thermal Hall conductivity $\kappa_{xy}$. In particular, 
in the Majorana insulator in two dimensions, the 
ratio $\frac{\kappa_{xy}}{T}$ approaches quantized values as the temperature goes to zero. Phases with 
different values of the quantized thermal Hall conductance are {\em topologically} distinct and are separated by 
phase transitions. In recent work with X.G. Wen\cite{pairqh}, we studied the physics of this novel quantum 
Hall system in some detail.

From the point of view of the replica non-linear sigma model field theory
discussed earlier in this section, the existence of insulating phases with quantized thermal Hall
conductance can be attributed to the presence of a topological term in the action. This takes the 
form
\be
\label{stop}
{\cal S}_{top} = \int d^2 x \frac{\theta}{16 \pi} Tr \lrb{Q\pd{x}Q \pd{y}Q}  .
\ee
The quantity multiplying $\theta$ is equal to $i \times integer$ for any given configuration
of the $Q$ field. Thus the partition function is periodic under $\theta \ra \theta + 2\pi m$ for any
integer $m$. Physically, just as in the Pruisken field theory for the conventional integer
quantum Hall transition, $\theta$ is proportional to the (bare) value of the $f$-particle
Hall conductivity $\sigma^f_{xy}$. As shown in Ref. \cite{pairqh}, this is in turn proportional to 
the ratio of the thermal Hall conductivity to the temperature. 

\section{Phase diagram and transitions}
We now discuss the (zero temperature) phase diagram and phase transitions for the dirty superconductor
with non-conserved spin. Again, we consider the cases with and without ${\cal T}$ separately.

\subsection{Time reversal invariant systems}
As argued in the previous section, ``metallic" and ``insulating" phases are 
possible in both $d=2$ and $d=3$. A schematic phase diagram as a function of the bare 
thermal conductivity is as shown below in Figure \ref{major3d}.

\begin{figure}
\epsfxsize=3.3in
\centerline{\epsffile{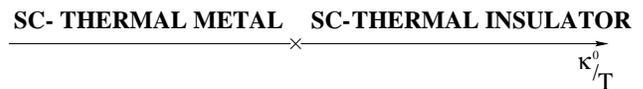}}
\vspace{0.15in}
\caption{Schematic zero temperature phase diagram for the three dimensional
superconductor in the presence of spin orbit scattering.}
\vspace{0.15in}
\label{major3d}
\end{figure}

The critical points separating the two phases are in new universality
classes for Anderson localization.  There are some interesting differences between the 
scaling properties of the two and three dimensional systems. 
In three dimensions, the ratio $\frac{\kappa}{T}$,
the thermal diffusion coefficient, $D$,  and the density of states $\rho_0$ at the 
Fermi energy are finite constants in the ``metal" and are zero in the ``insulator". 
It is therefore natural to expect that these will go to zero continuously (with some 
universal critical exponent) as the transition is approached 
from the ``metallic" side. In two dimensions however, in the ``metal", these quantities
are all infinite in the limit of zero temperature and infinite system size. 
(They are zero in the ``insulator"). Exactly at the transition point, conventional 
scaling arguments\cite{mpaf} imply a constant value for $\frac{\kappa}{T}$. The 
behaviour of the density of states 
and diffusion coefficient is more unclear. It is perhaps natural to 
suggest that these will be a constant, though we certainly cannot rule out other possibilities. 

\subsection{Time reversal non-invariant systems}
We first consider the three dimensional case. The phase diagram is similar to that above
for the time reversal invariant system. However, as the density of states may be 
non-zero in the ``insulator", we expect that it generically is non-zero at the transition point as well.
 
The two dimensional case has a richer phase diagram due to the possibility of
quantum Hall phases. 
For simplicity, we show, in Figure \ref{major2d}, only the two insulating phases with $\kappa_{xy} = 0$ and
$\frac{\kappa_{xy}}{T} = \frac{1}{2}$ (in units of $\frac{\pi^2 k_B^2}{3h}$). 
Note that there are three distinct phase
transitions, and a multicritical point in this phase diagram. 
First consider the Majorana metal-insulator transition.  
Both $\frac{\kappa_{xx}}{T}$ and $\rho_0$ are infinite in the Majorana metal. 
In the Majorana insulator, $\frac{\kappa_{xx}}{T}$ is zero while $\rho_0$ may be 
non-zero. As in the ${\cal T}$ invariant case, conventional scaling arguments 
imply a finite value for $\frac{\kappa_{xx}}{T}$ at the transition point. For the density of states, it is
natural to guess that it is finite at the transition, though again it is hard to rule out 
other possibilities. 

Now consider the transition from the Majorana metal to the quantum Hall phase. This transition
is in the same universality class as the transition from the Majorana metal to the insulator
with $\kappa_{xy} = 0$. The quantum Hall phase 
is a Majorana insulator with $\frac{\kappa_{xy}}{T} = \frac{1}{2}$. The distinction between these 
two insulating phases is due to the presence of (heat) current carrying edge states - these are expected to 
be unimportant in determining the properties of the transition to the metal. Below, we will provide a more
formal argument in support of this claim. 

Finally, consider the transition between the insulator with $\kappa_{xy} = 0$ and the one with 
$\frac{\kappa_{xy}}{T} = \frac{1}{2}$. A theory for this transition is obtained by 
considering a particular realization of the two phases. Consider a system of spinless fermions paired into
a $p_x +ip_y$ superconducting state in two dimensions. In Ref. \cite{pairqh}, it was shown that 
such a superconductor has a quantized thermal Hall conductivity $\frac{\kappa_{xy}}{T} = \frac{1}{2}$
if the chemical potential is positive (the ``BCS" limit), and has $\kappa_{xy} = 0$ if 
the chemical potential is negative (``molecular" limit). Thus, as the chemical potential is varied through
zero, there is a transition from the Majorana quantum hall phase to the Majorana insulator. Ref. \cite{pairqh}
also examined the theory for this transition, and argued that, at least at weak disorder, it is 
correctly described by a theory of relativistic Majorana fermions with random mass.
It is well-known that the random mass is irrelevant at the pure free Majorana fixed point\cite{RBIM}. 
Thus the critical
theory is known {\em exactly} in this case. If we make the important assumption
that there is a unique fixed point describing the Majorana insulator-quantum Hall
transition, then the arguments above identify it with the free relativistic Majorana fixed point.
There is, however, some reason to question this assumption - see Section \ref{Nish}.   

Some more insight into the phase diagram and the transitions comes from considering the 
properties of the replica non-linear sigma model field theory describing the quasiparticles in the two
dimensional superconductor with spin orbit scattering, and no ${\cal T}$. This is 
described by the action Eqn. \ref{sd} supplemented with the topological term Eqn. \ref{stop}. The 
small $g$ regime is described by the perturbative calculations of the previous section. 
Note that the presence of the 
topological term does not affect those results. Indeed, the parameter $\theta$ plays no role 
in perturbation theory, and does not renormalize. Thus there is a line of fixed points in 
the $g,\theta$ plane  at $g = 0$ with $\theta$ arbitrary. As the value of $\theta$ is 
proportional to the value of $\frac{\kappa_{xy}}{T}$, this ratio varies continuously in the Majorana metal.

Now consider the behaviour of the sigma model at large $g$. It is natural to 
expect that, in this limit, the physics is captured by a strong-coupling expansion, 
and that the resulting phase corresponds, physically, to a localized phase. (For conceptual
purposes, it may be convenient to think in terms of a lattice regularized version 
of the field theory.)
First consider $\theta = 0$. The corresponding localized phase has $\kappa_{xy} = 0$. 
Similarly, the insulator with $\theta = 2\pi$ has $\frac{\kappa_{xy}}{T} = \frac{1}{2}$. 
Just as in other sigma model field theories with topological terms, 
the large $g$ phase for $0 \leq \theta < \pi$ is expected to be continuously connected
to the large $g$ phase at $\theta = 0$, {\em i.e} it is an ``insulator"
with $\kappa_{xy} = 0$. Similarly the large $g$ phase with $\pi < \theta \leq 2\pi$
is expected to be continuously connected to the large $g$ phase with 
$\theta = 2\pi$, {\em i.e} an ``insulator" with  $\frac{\kappa_{xy}}{T} = \frac{1}{2}$.
The transition between the two localized phases occurs at $\theta = \pi$. 
The small $g$ ``metallic" phase is separated from these localized phases 
at large $g$ by a phase boundary. The symmetry $\theta \ra 2\pi - \theta$ implies that 
this phase boundary be symmetric about the $\theta = \pi$ line.

\begin{figure}
\epsfxsize=3.3in
\centerline{\epsffile{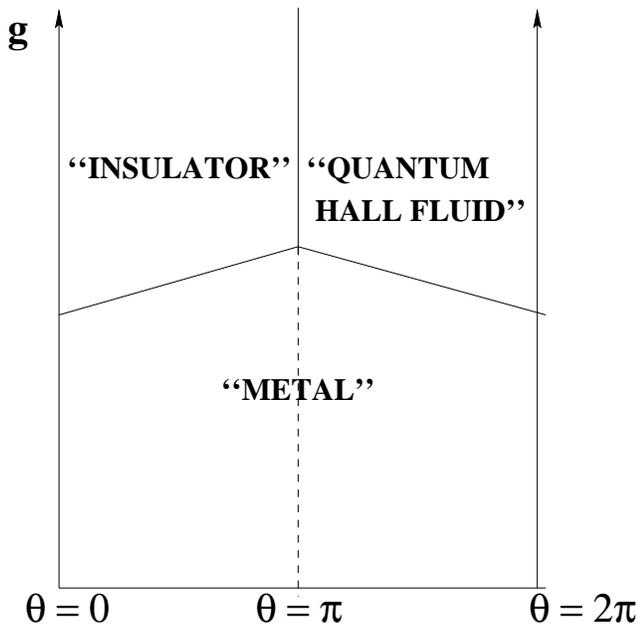}}
\vspace{0.15in}
\caption{ Phase diagram of the sigma model field theory 
describing the two dimensional superconductor with no spin rotation
or time reversal invariances.}
\vspace{0.15in}
\label{scsonotft}
\end{figure}   

In the field theory, the transitions between the small $g$ ``metal" and the large $g$
``insulator" at $\theta = 0, 2\pi$ are in the same universality class as the 
physics is invariant under $\theta \ra \theta + 2\pi$. What about the transition
from the ``metal" to the ``insulator" at other values of $\theta$ ? We suggest that,
for $\theta \neq \pi$, these are in the same universality class as the transition at $\theta = 0$. 
This then implies that $\kappa_{xy}$ is continuous across the ``metal-insulator" transition.
Finally, the point where the ``metal-insulator" phase boundary 
crosses the $\theta = \pi$ line in the phase diagram of the 
field theory corresponds to the multicritical point where the 
``metal" and the two ``insulating" phases come together.

\section{Random bond Ising model in $d = 2$}
\label{Nish}
In this section, we briefly discuss the possible implications of this paper to the finite
temperature properties of
classical random bond Ising models in two dimensions. It is well-known\cite{RBIM} that this system
admits a description in terms of non-interacting disordered Majorana fermions. 
Specifically, the partition function for the Ising model can be written as 
a functional integral over Majorana fields that live in two dimensions. This, in turn, 
may be reinterpreted as a generating function for the wavefunctions of a quadratic
quantum Hamiltonian of Majorana fermions in two spatial dimensions. This Majorana
Hamiltonian has no special symmetries - the system is therefore in the universality
class of the two dimensional superconductor with neither spin rotation nor 
time reversal invariance.

It is therefore natural to try to identify
various phases of the Ising model with corresponding ($T = 0$) phases in the two dimensional 
superconductor with no spin rotation or time reversal invariance. First consider the 
pure Ising model. Both the high and low temperature phases correspond to gapped phases
in terms of the Majorana fermions. Thus these correspond to insulating phases of the 
Majorana fermions. Disorder in the Ising model, obtained by making some of the bonds
random, would tend to fill up the gap. However, if the disorder is weak, the resulting low 
energy states will be strongly localized. Thus, at weak disorder, both the high and low temperature
phases of the Ising model correspond to localized phases of the Majorana fermions.
How then do we distinguish between the two? The distinction is topological, with 
the quantized value of the thermal Hall conductance differing by $1/2$ between the two phases.  
Strong support for this suggestion is obtained by examing the properties of the system near 
the transition
between the two phases. In the pure Ising model, the critical point is described by a 
massless relativistic Majorana theory. Moving off criticality introduces a mass
for the fermions with the sign of the mass distinguishing the two phases. 
The result that the dimensionless ${\kappa}_{xy}$ jumps by $1/2$ at the transition
can now be established by direct calculation. The critical theory is identical to the 
one describing the 
transition, at weak disorder, in the $p_x +i p_y$ superconductor of spinless fermions between the 
BCS and molecular limits. We thus identify the Ising transition at weak disorder with the 
thermal quantum Hall transition.

\begin{figure}
\epsfxsize=3.5in
\centerline{\epsffile{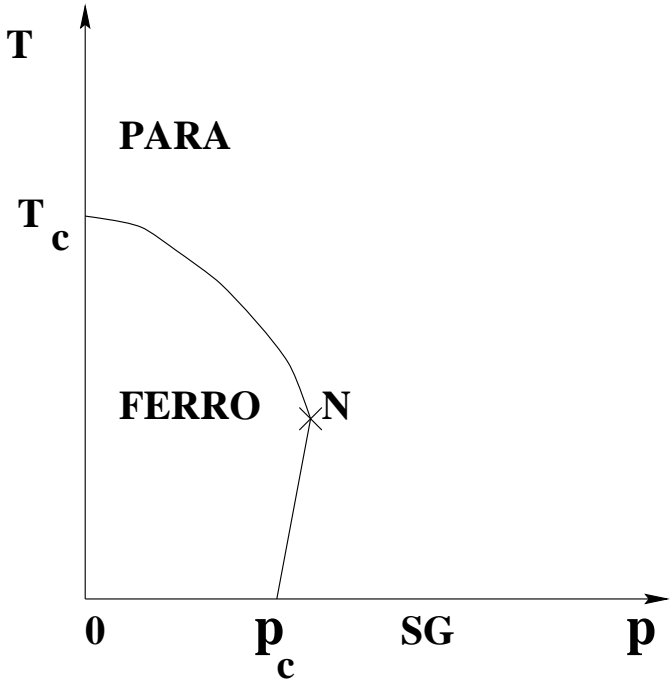}}
\vspace{0.15in}
\caption{ Phase diagram of the two dimensional Ising model with random bond strengths. $p$ is 
the concentration of negative bonds, and $T$ is the temperature. The point $N$ is the Nishimori
multicritical point. SG refers to the spin glass phase which is 
believed to exist only at zero temperature.}
\vspace{0.15in}
\label{Nishf}
\end{figure}

Now consider introducing strong (bond) randomness into the Ising model. If all the bonds
are still positive, the transition is believed to be always controlled by the pure Ising fixed point.
The situation is more interesting if some of the bonds are made negative. As the concentration of 
negative bonds is increased, there is a multicritical point (the 
Nishimori point, see Figure \ref{Nishf})\cite{Nishi} on the phase boundary, beyond which the 
transition is no longer controlled by the pure Ising fixed point. 
The properties of this multicritical point have been the focus of a number of
investigations\cite{Nishi1} over the last many years. Despite this, there is no detailed understanding
of the theory of this point, or of the properties at the phase boundary at lower temperatures.
What does the 
the Nishimori point correspond to in the language of the Majorana fermions? 
There appear to
be two distinct possibilities - we will discuss evidence supporting either below.

(i) For the problem of localization of Majorana fermions in two dimensions,  
the assumption that there is a unique fixed point controlling the thermal quantum 
Hall transition leads to it's identification with the free relativistic Majorana theory. 
This assumption, which is perhaps natural in the fermion language, then points to the identification
of the Nishimori point with the multicritical point separating the Majorana metal, insulator, 
and the quantum Hall 
phase. If true, then it is natural to expect that the ferromagnetic phase transition in the Ising
model at low temperatures below the Nishimori point is actually the Majorana insulator-metal transition.
Some evidence in support of this is provided by the numerical results of Ref. \cite{sora} which
found no signs of localization of the fermions in the non-ferromagnetic phase at low temperature 
close to the phase boundary. On the other hand, note that the paramagnetic 
phase at high temperature and weak disorder corresponds to a localized phase. If the 
scenario outlined above is correct, then we are led to infer the existence of a 
finite temperature phase transition outside the ferromagnetic phase in the Ising model associated 
with a delocalization of the Majorana fermions. It is unclear what this 
transition means in the Ising language, and even whether it happens at all. 
A natural candidate would have been
a transition to a spin glass phase at low temperature - however there is strong numerical evidence 
for the absence of spin glass order at finite temperature in two dimensional Ising systems.  It seems possible
that a delocalized (``metallic") phase of Majorana fermions
would correspond to
a phase in which both the Ising spin
and the dual Ising disorder parameter are simultaneously zero.

(ii) A different scenario is obtained by dropping the assumption that there is a unique 
fixed point controlling the thermal quantum Hall transition. Instead, for the Ising
model, we assume that the only finite temperature phase transition is associated with the 
destruction of the ferromagnetic order. If the concentration of negative bonds is large enough
to destroy the ferromagnetism at zero temperature, we assume that the resultant state
is a spin glass. These assumptions are perhaps reasonable expectations for the Ising model. 
Then, the finite temperature ferromagnetic and paramagnetic phases are, in the fermion language both
localized phases which are distinguished by their thermal Hall conductance. The existence of the Nishimori
point then implies the existence of a multicritical point (at strong disorder) in the phase boundary
between these two localized phases. Then, the thermal quantum Hall transition 
will be controlled by the free
relativistic massless Majorana theory at weak disorder, but by a different fixed point at strong disorder.
The spin glass phase in the Ising model presumably corresponds to the Majorana metal. However, it may seem
a bit puzzling, in this scenario, why the Majorana metal which is a stable phase of disordered
Majorana fermions in two dimensions is only realized in a set of measure zero (a line at $T = 0$)
in the Ising phase diagram.

Which one of these two scenarios is actually realized and the resultant consequences both for the  
Ising model and the Majorana localization problem, we leave as an intriguing open question.

\section{Discussion}
In this paper, we have discussed the physics of localization of quasiparticles 
in a superconductor in situations where the quasiparticle spin is not a good quantum number.
Our discussion was entirely based on models of non-interacting quasiparticles - including
the effects of interactions is an important and interesting issue which we leave for future work.
Even within the non-interacting theory, we have found a rather rich phase diagram, and a number of 
as yet unexplored phase transitions. 
Several experimental systems to which this work is of relevance 
can be imagined - here we briefly mention a few. 

(i) A particularly attractive prospect for probing the physics discussed here
is in a Type $II$ superconductor in a strong magnetic field in the presence of
spin-orbit scattering impurities. At low fields, the system will be in the superconducting
``insulator" phase. With increasing field, under conditions discussed in Ref. \cite{smitha},
there will be a transition (at zero temperature) to a superconducting ``metal" phase. 
This transition can be
probed, for instance, by measurements of the low temperature quasiparticle
heat conductivity\cite{smitha}. 

(ii) A number of recent experiments\cite{Taill} have measured low temperature heat transport 
by the quasiparticles in the heavy fermion superconductors. We note that as these
systems typically have strong spin-orbit scattering, the results of this paper are of 
potential importance. Some of the heat conductivity measurements have been motivated 
by the possibility of identifying the correct pairing symmetry in these superconductors. 
Such identification could be seriously hampered by the localization issues discussed here.
In particular, if the disorder is strong enough to localize the quasiparticles, it is 
hard to infer whether the pure system has a gap to quasiparticle excitations or not from
heat transport measurements alone.

(iii) The properties of superfluid $He-3$ in porous media have been the subject of 
some experimental studies\cite{parpia}. In this context, it is interesting to ask if the fermionic
$He-3$ quasiparticles in the superfluid are localized or delocalized 
at zero temperature. This may, perhaps,
again be probed by heat transport experiments. It also interesting to consider
the properties of superfluid $He-3$ on a disordered two dimensional substrate. 
In the $A$ phase, at weak disorder, a thermal quantum Hall effect\cite{pairqh}
is predicted.  
With increasing disorder, it is possible (though not necessary) that there is a zero temperature
phase transition where the quantization of the thermal Hall conductivity is destroyed {\em before}
the superfluidity is destroyed. If this happens, this would be an experimental realization 
of the thermal quantum Hall transition discussed in the previous section\cite{pairqh}.

Finally, the general problem of localization of Majorana fermions arises in other contexts
as well - a specific example being the random bond Ising model in two dimensions as 
discussed briefly in Section \ref{Nish}. Progress in the localization problem may 
therefore provide a route to further our understanding of the random bond Ising model.

We thank I.A. Gruzberg and Martin Zirnbauer for useful discussions.
This research was supported by NSF Grants DMR-97-04005,
DMR95-28578
and PHY94-07194.

\end{multicols}
\end{document}